# Concomitant opening of a topological bulk-gap with an emerging Majorana edge-state

Anna Grivnin, Ella Bor, Moty Heiblum, Yuval Oreg, and Hadas Shtrikman

*Braun Center for Submicron Research, Weizmann Institute of Science, Rehovot 76100, Israel*

**Majorana 'zero-modes' are expected to be immune to decoherence. The primary method for their characterization in a 1D topological superconductor, is measuring the tunneling current into the edge of the superconductor. Presently, the hallmark of a localized Majorana edge-state is an emergent quantized zero-bias conductance peak (ZBCP). However, such a conductance peak can also result from other mechanisms; *e.g.*, crossing (and sticking) of two branches of a standard Andreev bound state, or a soft potential at the edge of the superconductor. Since the emergence of a 'Majorana-ZBCP' must be accompanied by an opening of a topological gap in the bulk, we performed two simultaneous tunneling measurements: one in the bulk and another at the edge of the 1D superconductor. Measurements were performed with an InAs nanowire in-situ coated with epitaxial aluminum. For particular gate-tuning of the chemical potential in the wire and a Zeeman field parallel to the wire, we observed a closing of the superconducting bulk-gap followed by its reopening concomitant with the appearance of a ZBCP at the edge. We note that a ZBCP could also be observed with different tuning parameters without an observed reopening of the bulk-gap. This demonstrates the importance of simultaneous probing of the bulk and edge when searching for a Majorana edge-state.**

## Introduction

The proposed prescription for the formation of a 1D topological superconducting phase [1-10], ignited a spree of experimental works [11-22]. With a 1D-like semiconducting nanowire (from now, 'wire') that bears spin-orbit interaction, applying a parallel (to the wire) Zeeman field, helical modes are expected to emerge in a restricted range of the chemical potential in the wire. Proximitizing the wire to a 'trivial' (s-wave) superconductor is expected to induce a topological superconducting gap, accompanied by a pair of localized 'Majorana zero-energy modes' at the two opposite ends of the wire. Consequently, tunneling electrons into the edge of the wire is expected to feature a 'zero-temperature' quantized zero-bias conductance peak (ZBCP), $G_M=2e^2/h$ (e – electron charge, h – Planck's constant). Such ZBCPs had been observed in previous works with proximitized InAs and InSb 1D-like wires [11-20]. A recent observation of a quantized value of $G_M$ was reported in Ref. 22; for a review see Ref. 21. A ZBCP was also reported in tunneling experiments in ferromagnetic 1D atomic chains [23-25]. However, the most crucial ingredients for a topological superconductor, the presence of helical modes and the ability to 'gate-control' the chemical potential, were, to the best of our knowledge, not verified. Moreover, since a 'nearly quantized' ZBCP can also emerge at the edge of a 1D 'trivial superconductor' due to alternative mechanisms [*e.g.*, 26-29], a direct evidence of a Majorana edge state is warranted. Supporting evidence is an observation of a bulk-gap opening with a concomitant appearance of a ZBCP at the edge of the wire [30].

Kitaev provided a general procedure for the emergence of Majorana edge states in a topological 1D superconductor [1]. Such a 1D topological superconductor can be induced by coupling a trivial superconductor to a semiconducting 1D-like wire that harbors Rashba spin–orbit coupling, accompanied by a parallel Zeeman field to the wire [9, 10]. A sharp phase transition, which separates the topological and the trivial superconducting phases is expected to take place at $E_z = 2\sqrt{\Delta_{ind}^2 + \mu^2}$, with $\Delta_{ind}$ the superconducting gap in the wire, $E_Z = g\,\mu_B B$ the Zeeman splitting, $g$ the Landé $g$ factor and $\mu_B$ the Bhor magneton. The condition $\mu = 0$ is dubbed as the 'sweet spot' – when the Fermi energy is located at the crossing energy of the two spin--orbit-split energy dispersions [31-35]. A further increase in $E_z$ reopens a (p-wave) topological gap. When the Fermi energy is within the Zeeman gap, the effective superconducting gap is the smallest of the two gaps; one at $k = 0$ and the other at $k = k_F$. For a small field the dominant bulk-gap is $E_g = 2\left|E_Z/2 - \sqrt{\Delta_{ind}^2 + \mu^2}\right|$ at *k*=0. At a high field the dominant gap is at $k = k_F$, decreasing with an increasing Zeeman field. The reopening of a topological gap must be accompanied by a zero-energy bound state at the end of the wire – the localized Majorana zero mode.

In the following, we describe our approach of simultaneous measurements of the energy dependent tunneling density of states with a few tunneling probes placed at the bulk and at the edge of the proximized wire.

## Experimental set-up

A scanning electron micrograph of the device and its schematic illustration are shown in Figs. 1a, 1b and 1c. Stacking faults free Wurtzite InAs wires, with 50–60nm diameter, were grown by Au-assisted vapor-liquid-solid process in a molecular beam epitaxy (MBE) system on a (001) InAs substrate. For the wires growth, a very thin gold layer was evaporated on the InAs substrate after oxide blow-off in a separate vacuum chamber connected to the MBE system. After the wires' growth, a ~7nm thick Al layer was in-situ evaporated on the side of the wires at a substrate temperature of -40C, creating a half-shell. The wires were spread on a Si/SiO$_2$ substrate, already pre-covered by a wide metallic gate and 50nm thick HfO$_2$ (by ALD). Tunneling probes (TPs), 200nm wide, were made with a ~1nm thick aluminum-oxide barrier and a top TiAu contact. A few TPs were placed at the bulk region and one TP at the edge of some 6µm long proximitized wire(s) [36-39]. The epitaxial Al was contacted by a TiAl superconducting contact (some 2.8µm from the edge). As shown in Fig. 1c, the TPs coupled only to the InAs wire (as the epitaxial Al was oxidized). It is worth stressing that in contrast to most pervious configurations, with bare regions along the wire, our wires were *fully* lengthwise covered by the Al superconductor. This allowed to mitigate undesirable gate-induced potential in the bare parts of the wire (*e.g.*, creating unintentional quantum dots, locally formed Andreev bound states, etc.).

Measurements were performed in a dilution refrigerator at 14mK. The 'bulk' and 'edge' TPs were excited by $V_{AC}$=2µV RMS on top of a variable DC bias, $V_{DC}$. Each TP was excited by a different signal frequency, allowing two simultaneous (lock-in-amplifier type) measurements (Fig. 1a). The induced superconducting gap in the wire was found to be close to that of the intrinsic Al film, $\Delta$~250µeV (Fig. 1d). The critical parallel to the wire magnetic field (accuracy ±2 degrees) was ~1.6T.

## Results

Spectroscopy, in the form of tunneling conductance $dI/dV$ as function of DC bias $V_{DC}$ and magnetic field $B$, is shown in Fig. 2. Measurements with the edge-TP (TP1, tunneling resistance ~100KΩ) and the two bulk-TPs (TP2, ~83kΩ and TP3, ~92kΩ), was performed at a back-gate voltage $V_{BG}$=-4.22V. This range of the tunnel resistance was found to be optimal (after multiple trials), as it allowed a measurable tunneling current while still minimizing field penetration into the wire. We believe that the potential fluctuations in the wire are stronger than the induced potential by the TPs.

Figure 2a shows the evolution of the tunneling current at the edge of the wire as function of the Zeeman field and applied DC bias (TP1, colored yellow in the inset). Two branches of an Andreev bound state merge at zero bias to a single ZBCP at $B$~0.7T, which persists up to $B$~1.6T (when the Al superconductivity quenches). The ZBCP height is $G_M$=0.22$e^2/h$ - likely limited by the finite temperature, the weak tunneling coupling, and the finite dissipation [40]. The simultaneously measured evolution of the bulk-gap with TP2 (positioned ~500nm away from the edge, colored yellow in the inset), is shown in Fig. 2b. The gap closes at $B$~0.7T and reopens at $B$~1.05T - increasing to $E_g$~65µeV at $B$~1.4T. Notice that our resolution was ~25µeV, thus limiting an exact identification of gap closure / opening. Also, notice, that finite size effects may also lead to a lower bound of the bulk gap [45]; but in a 'micron-long' topological region this bound is below our resolution limit. A similar behavior is observed with TP3 in Fig. 2c (placed ~1200nm from the edge, colored yellow in the inset). Here, the gap closes at $B$~0.95T and reopens to $E_g$~30µeV at $B$~1.4T. This difference is likely due to a variation in the local chemical potential along the wire - highlighting the difficulties in measuring nanowires [41-44]. Notice that the ZBCP reaches a maximal height when the topological gap is the largest (see Fig. 1 in Extended Data).

We studied the evolution of the gap as function of the back-gate voltage (Fig. 3). As, due to the small mutual capacitance between the TP and the wire, the gate bias is not expected to affect the tunneling probes significantly, it is fair to assume that the observed dependence on the gate voltage is due to a change in the chemical potential in the wire. Starting with Fig. 3a, where $V_{BG}$=-4.22V (same as in Fig. 2), the trivial bulk-gap closes at $B$~0.7T and reopens at $B$~1.05T, to finally saturate at $E_g$~65µeV (see Fig. 1 in Extended Data). Being the largest observed gap, it is likely that the Fermi energy is placed very close to the 'sweet spot'. Increasing the back-gate voltage slightly to $V_{BG}$=-4.21V, the transition field moves up to $B$~1T and the gap saturates at $E_g$~25µeV (Fig. 3b). At $V_{BG}$=-4.13V the gap closes at $B$~1.1T, and never reopens in the full range of magnetic field (Fig. 3c). One possible scenario (among others [29, 46]), is that the chemical potential is too high, thus, an opening of a topological gap requires a Zeeman field that is near the critical field. At this chemical potential $g^*$=3.9, being the largest we measured in this device. A strong correlation of the appearance of the ZBCP with TP1 with the reopening of the bulk-gap is shown in Figs. 3d and 3e. Alternatively, while the bulk-gap does not reopen in Fig. 3f, a 'trivial' ZBCP appears. These findings clearly demonstrate the advantage of simultaneous probing of the bulk and the edge for the identification of a topological state.

The temperature dependences of the ZBCP (Fig. 4a) and bulk-gap (Fig. 4b), at the 'sweet spot' of the chemical potential and $B$=1.25T, was measured in the range $T$=14-200mK. At base temperature, the full-width at half-maximum of the ZBCP was ~15μeV, which is in reasonable agreement with the combined effect of thermal broadening ($3.5k_BT$~4.5μeV at 14mK), the excitation voltage (5.6μeV p-p), and the tunneling broadening. The evolution of the ZBCP is plotted as function of temperature in Fig. 4c. The expected height was also calculated by a convolution of the derivative of the Fermi function and the conduction trace d$I$/d$V$ at base temperature (which already contained the temperature independent broadening due to lifetime and excitation signal). The bulk-gap starts at $E_g$=65μeV at $T$=14mk and softens with temperature (Fig. 4b). As expected, a complete gap softening, due to thermal broadening (~$3.5k_BT$), is observed around $T$=100mK.

## Summary

The local density of states was measured via tunneling spectroscopy at different positions along a fully proximitized InAs nanowire with a trivial superconductor (without bare regions along the wire). Tunneling probes were placed on the wire at the 'bulk' and at the 'edge'. Under a specifically tuned Zeeman field and chemical potential, correlation between bulk-gap closure (of the 'trivial gap') followed by its reopening (as a 'topological gap'), and the emergence of a zero-bias-conductance-peak (ZBCP) at the edge was observed. This is expected for an emergence of a localized Majorana state at the edge of the nanowire. The dependence of the 'bulk-gap' and the ZBCP on the back-gate voltage demonstrates the importance of correct tuning of the chemical potential in the bulk. We also find that for certain values of the chemical potential and magnetic field a ZBCP can appear at the edge without evidence of reopening of a gap in the bulk. A simultaneous study of the bulk and the edge of the wire may add confidence and serve as a supporting component, in the identification of a localized Majorana zero mode.

**Acknowledgements**

We thank H. Inoue, A. Haim, Y. Ronen, Y. Cohen and Y. Reiner for useful discussions. We are grateful to J.-H. Kang and D. Mahalu for their professional contribution, and to Michael Fourmansky for professional technical assistance. We thank S. Das Sarma and J. Alicea for their useful comments on the manuscript. M.H. acknowledges the partial support of the Israeli Science Foundation (ISF), the Minerva foundation, the European Research Council under the European Community's Seventh Framework Program (FP7/2007–2013)/ERC - Grant agreement 339070. H.S. acknowledges partial financial support of the Israeli Science Foundation (Grant No. 532/12 and Grant No. 3-6799), Israeli Ministry of Science (Grant No. 0321-4801 (16097)) and BSF grant No. 2014098. H. S., incumbent of the Henry and Gertrude F. Rothschild Research Fellow Chair. Y.O. acknowledges support by the BSF and ISF grants and by the European Research Council under the European Community's Seventh Framework Program (FP7/2007–2013)/ERC - Grant agreement MUNATOP-340210.

The data that support the plots within this paper and other findings of this study are available from the corresponding author upon reasonable request.

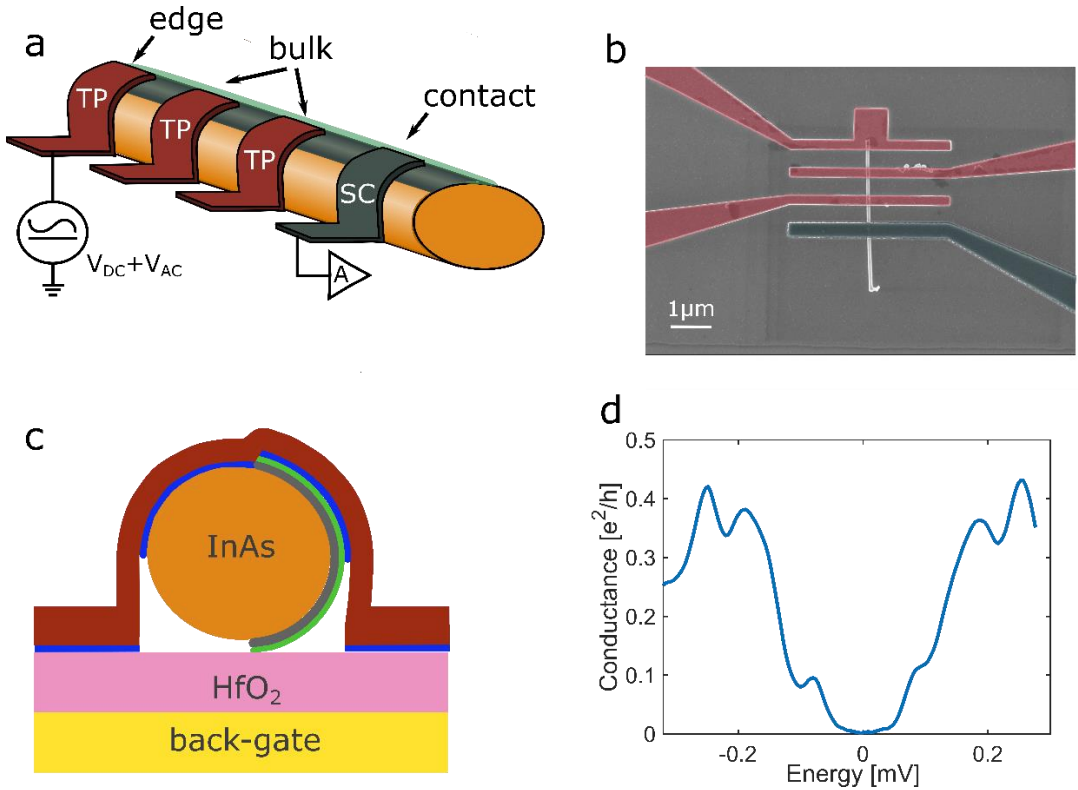

**Figure 1. The nanowire structure and its non-linear conductance.** a) Schematic illustration of the nanowire and tunnel probes (TPs). A superconducting Al contact, connected to the epitaxial Al on the wire (both grey). Several TPs are connected to the edge and bulk of the wire (red). Each TP tunnels only to the wire, since the epitaxial Al is isolated by a thick native Al oxide. b) An SEM micrograph of a typical device with three TPs (red) and a grounded superconducting contact (gray). c) Schematic side view of a TP region: The InAs wire (orange) was covered by an epitaxial layer of epitaxial Al (7nm thick, grey, on right), which is oxidized (oxide thickness is 3nm, green). The wire is placed on top of a metallic back-gate, which is covered by 50nm thick $HfO_2$ insulating layer. The TP is made of a thin Al-oxide barrier (blue) and a metallic Ti/Au contact (red). d) The typical differential conductance measured by the bulk-TP shows a hard-gap of $\Delta_{ind}$=250µeV (at *B*=0 and *T*=14mK).

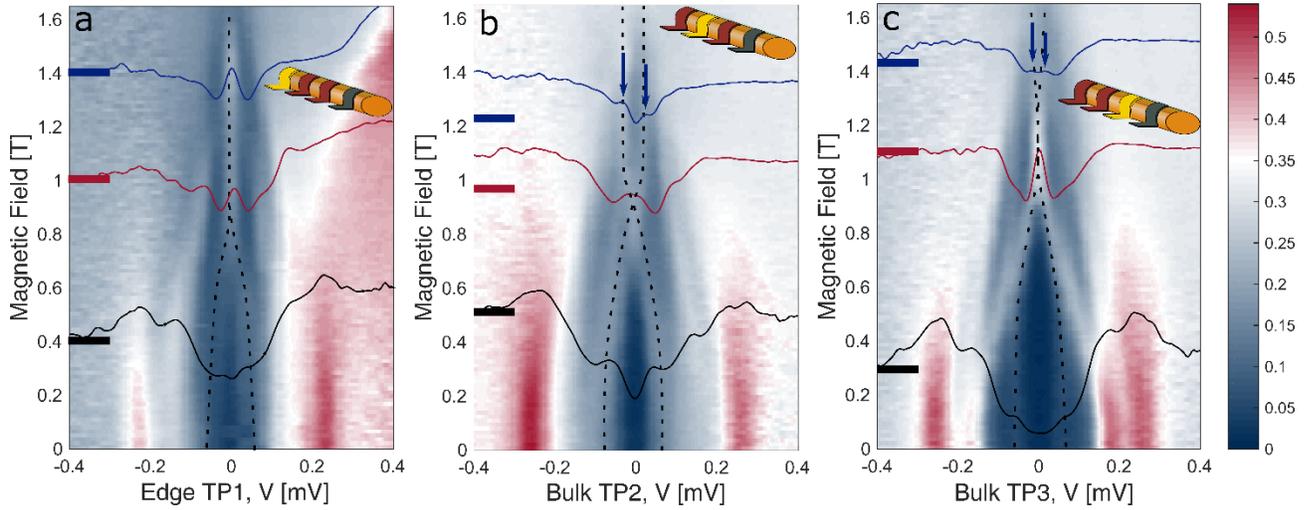

**Figure 2. Conductance spectroscopy at the edge and at the bulk of the nanowire.** Conductance as a function of the DC bias $V$ and the magnetic field $B$, measured by three tunnel probes: a) TP1 – edge; b) TP2 – bulk; c) TP3 – bulk. The active tunnel probe is colored in yellow in the inset of each plot. Cuts of the 3D plot at three different magnetic fields (marked by thick short arrows) are drawn on top (black, red, blue). a) At the edge a ZBCP appears at $B\sim 0.7$T and persists until 1.6T. b) In the bulk the trivial gap closes at $B\sim 0.7$T and reopens (as a topological gap) at $B\sim 1.05$T. The topological gap is marked by two blue arrows, with $E_g=65\mu eV$ at B=1.4T. c) The trivial gap closes at $B\sim 0.95$T and reopens at $B\sim 1.4$T, with $E_g=30\mu eV$. In all the measurements the back-gate voltage is $V_{BG}=-4.22$V. The dotted lines on all three figures are guides to the eye.

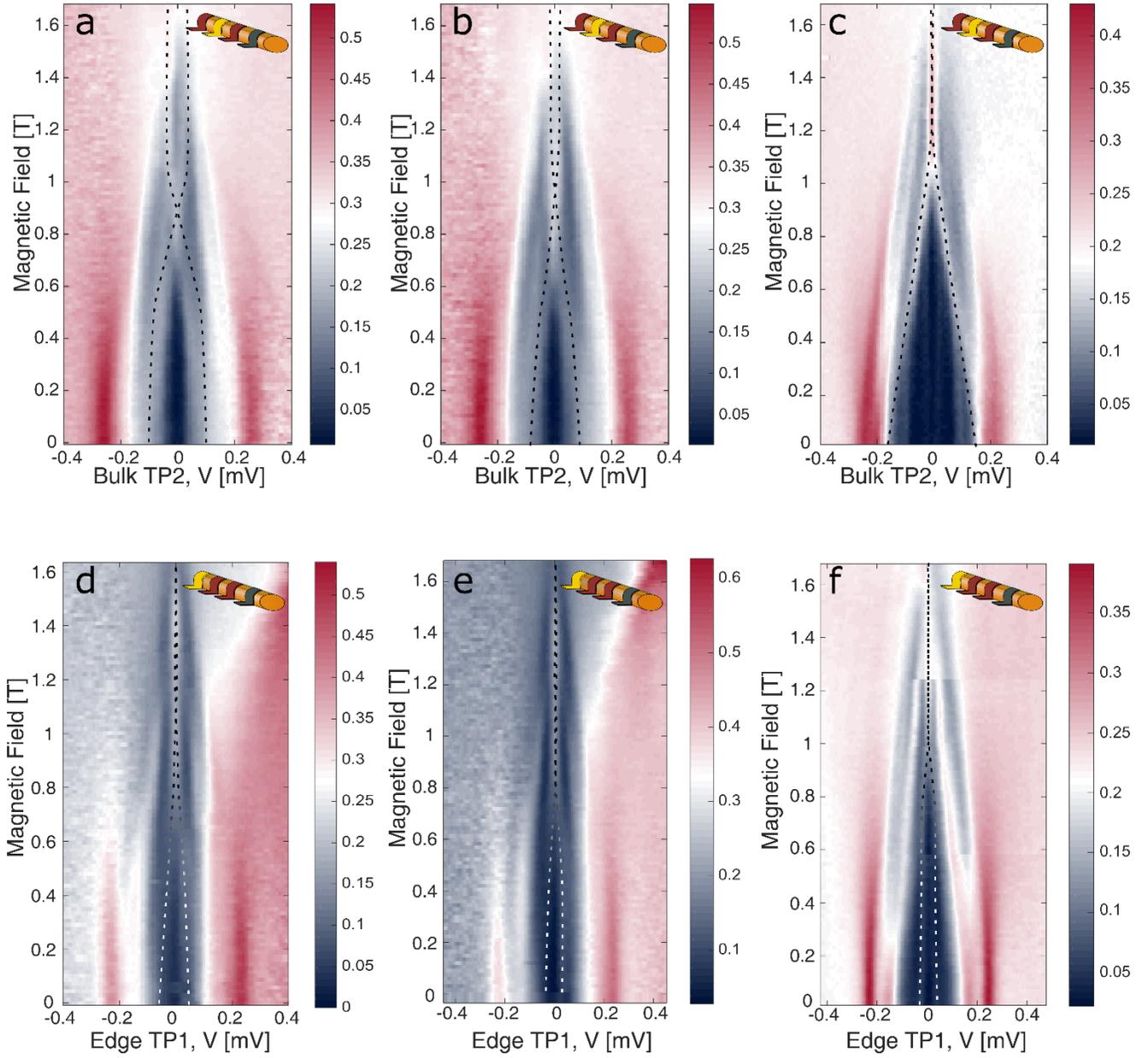

**Figure 3. Correlation between bulk gap closing and the appearance of a ZBCP.** Spectroscopy of the bulk (edge) as function of back-gate voltage (with TP2 (TP1)). The chemical potential is tuned by the back-gate voltage: $V_{BG}$=-4.22V (a, d), $V_{BG}$=-4.21V (b, e), $V_{BG}$=-4.13V (c, f). a) The chemical potential in the 'sweet spot' with bulk gap reopening at $B$~1.05T, reaching $E_g$~65µeV at $B$=1.4T. b) The chemical potential is slightly away

from the 'sweet spot', with topological $E_g$~25µeV. c) Gap closing without reopening. In all three values of $V_{BG}$ a ZBCP appears at the edge of the sample (see d-f). Only in two cases (a and b) the bulk gap reopens. This observation shows the importance of simultaneous measurements in the bulk and in the edge. The dotted lines are guides to the eye.

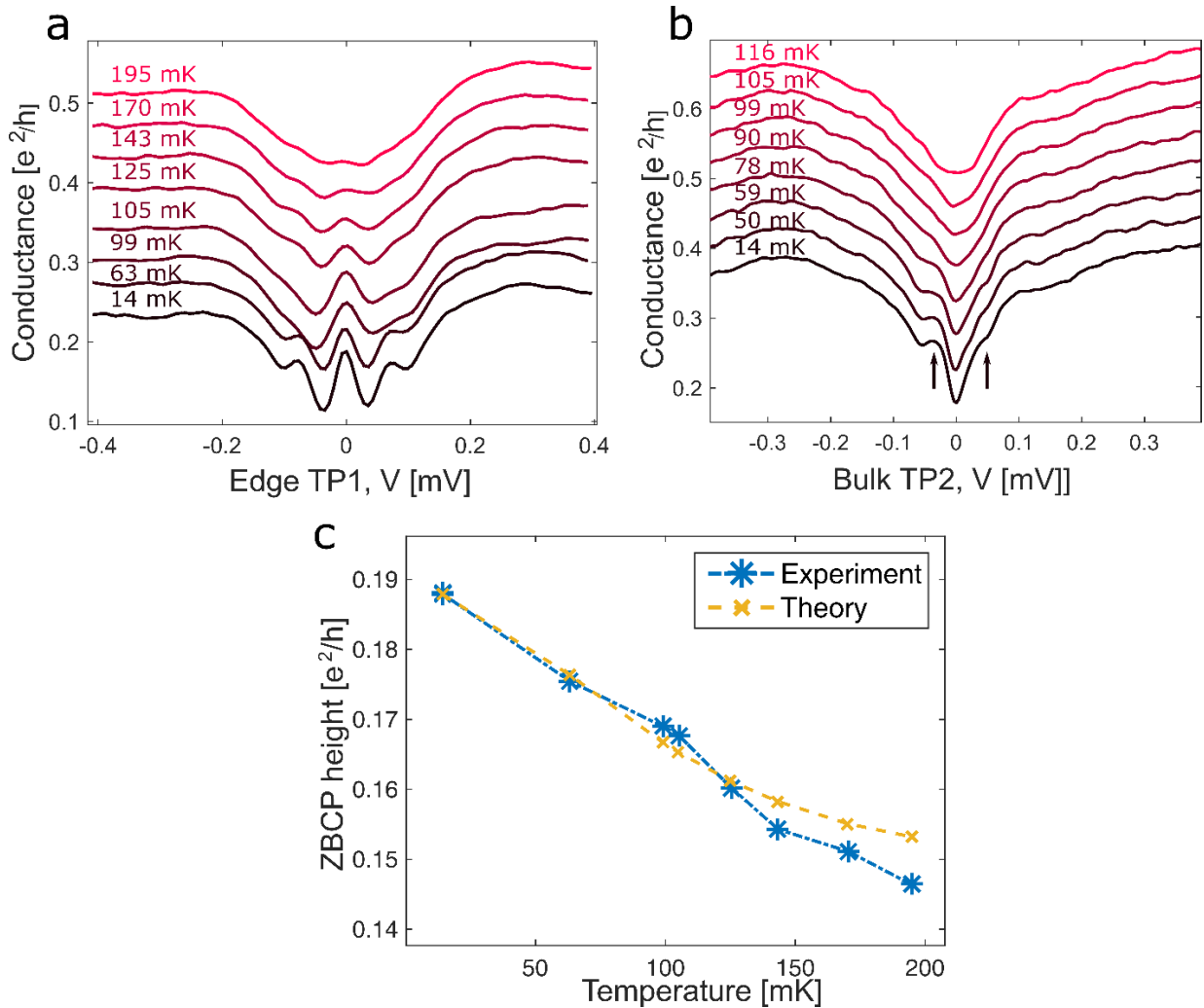

**Figure. 4. Temperature dependence of the ZBCP and the bulk-gap.** a) Dependence of the ZBCP on temperature. It is barely visible at $T$=~195mK. b) Dependence of the bulk-gap on temperature. At $T$~99mK the superconducting 'gap-shoulders' disappear. Plots are shifted by $0.04 \times e^2/h$ for clarity. c) The measured height of the ZBCP (blue) is plotted and is compared to a theoretical estimate (orange) - calculated by a convolution of the derivative of the Fermi distribution with the differential trace of the ZBCP at base temperature (the temperature independent broadening had been added).

**Methods**

*Sample fabrication*

For the deposition of the external superconducting contact to the Al thin layer on the wire, argon ion-milling process was used to remove the aluminum oxide (with beam voltage $V_B$=400V for $t$=4min), followed by in-situ evaporation of a Ti/Al (5/80 nm) contact. The tunnel probes (TPs) were fabricated via gentle ion milling on the InAs wire for oxide cleaning (with beam voltage $V_B$=400V for $t$ =1min) followed by a thin aluminum oxide barrier of ~1nm, which was formed in two steps: 1. Evaporation of Al at a rate of 1.7A/s for 7sec. 2. Putting the sample in the load-lock chamber of the evaporator, followed by an introduction of ozone at a pressure of 4.5x10$^2$ torr for 30 minutes. Afterwards, the sample was put back into the main chamber for evaporation of the metallic contact of Ti/Au (5/80nm). The TPs were 200-250nm wide. Their barrier is not likely to be affected by the back-gate voltage.

**An approximate calculation of important parameters of the device**

We assume a single band in the wire, with a 1D density of:

$$n = \frac{\sqrt{8m^*(\Delta_{so}+\Delta\mu)}}{\pi\hbar} \approx n_0\left(1 + \frac{1}{2}\frac{\Delta\mu}{\Delta_{so}}\right).$$

Here, $n$ is the electron density in the wire, $m^* = 0.023\,m_e$ is the effective band mass (with $m_e$ being the electron mass), $\Delta_{so} \sim 500\,\mu eV$ is the spin-orbit coupling energy[1], and $n_0 = \frac{\sqrt{8m^*\Delta_{so}}}{\pi\hbar} \approx 10\,\frac{\text{electrons}}{\mu m}$. The condition where $\Delta\mu = 0$ is referred to as the 'sweet spot'.

At the 'sweet spot', the bulk gap,

$$E_g = 2\left|E_Z/2 - \sqrt{\Delta_{ind}^2 + \Delta\mu^2}\right|,$$

closes at the lowest value of the Zeeman field, $E_z = g^*\mu_B B$, being equal to the induced superconducting gap in the wire $\Delta_{ind}$. From Fig 3a, corresponding to the 'sweet spot' (with $V_{BG}$=-4.22V), we find that the critical field for gap closure $B \approx 0.9\,T$ (this is an estimate, as the gap closes at $B \approx 0.7\,T$ and reopen at $B \approx 1.05\,T$), with the Zeeman energy $E_z \approx 210\,\mu eV$. At $V_{BG}$=-4.21 V the gap closes at $B \approx 1T$, and thus the change in the chemical potential is given by:

---

[1] The spin orbit energy $\Delta_{so}$ is by definition as the energy difference between the bottom of the band and crossing point of the shifted parabolas in the presence of spin-orbit coupling.

$$\Delta\mu = \sqrt{\frac{E_z(B=1T)^2}{4} - \Delta_{ind}(B=1T)^2} \approx 210\sqrt{\left(\frac{1}{.9}\right)^2 - 1} \approx 100\mu eV.$$

The *levering* factor is:

$$\frac{\Delta V_{BG}}{\Delta\mu} \approx \frac{10000}{100} \approx 100.$$

Hence, the change in the density of the electrons $\Delta n$, due to the change in the back-gate voltage $\Delta V_{BG} \approx 10000 \mu eV$, is:

$$\Delta n \approx n_0 \frac{1}{2}\frac{\Delta\mu}{\Delta_{so}} \approx 10\frac{100}{1000} \approx 1 \frac{electrons}{\mu m}.$$

The estimated capacitance between the back gate and the wire is:

$$C = \frac{e\,\Delta n}{\Delta V_{BG}} \approx 16\,\frac{pF}{m}.$$

We can also estimate the capacitance between the wire and the back-gate by using the classical formula, where *h*=75nm is the distance of the center of the wire from the back-gate, *r*=25nm is the radius of the wire, and $\varepsilon$ is the dielectric constant of HfO₂:

$$C_{classical} = \frac{2\pi\varepsilon\epsilon_0}{\cosh^{-1}(\frac{h}{r})} \approx 300\,\frac{pF}{m}.$$

The order of magnitude discrepancy between the experimentally measured capacitance and the classical value are probably due to the crude simplicity of the classical model, as it does not include screening effects and the change in the position of the electron wave-function inside the wire due to the presence of epitaxial Al. Also, the approximation of the classical capacitance assumes that the distance between the plane and the wire (*h*) is much bigger than *r*, which is incorrect for our case.

# Extended Data

Here, we present additional data, measured in other devices and on the device used in the main text after thermal recycling. In addition, we show the tunneling conductance in other regimes of the chemical potential and the magnetic field.

## ZBCP and bulk-gap as function of Zeeman field

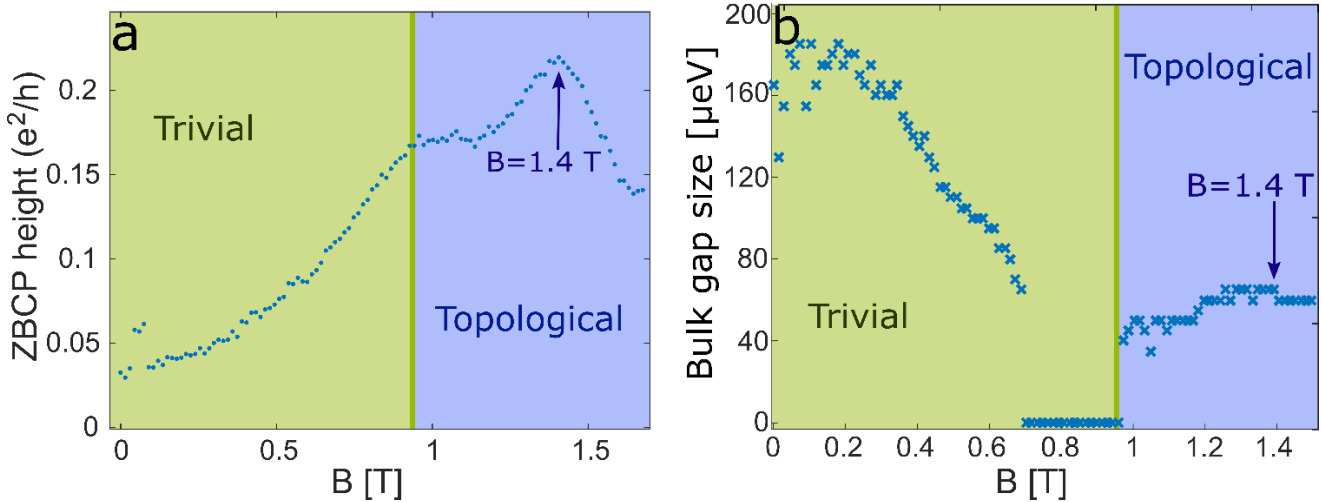

**Extended Data Figure 1.** a) ZBCP, b) bulk-gap – both function of the Zeeman field of the device presented in the main text. At the trivial region, $B<0.95T$, the gap decays with magnetic field while the ZBCP height rises. At the topological region, the ZBCP height increases with $B$, reaching a maximum where the topological gap is maximal ($B\sim1.4T$). For higher magnetic fields both ZBCP and the bulk-gap decay with $B$ due to the softening of the superconductivity. For magnetic fields higher than 1.5T the gap 'shoulders' smear away and the calculation of the gap size is unreliable.

**Measurement on a different device (device #2)**

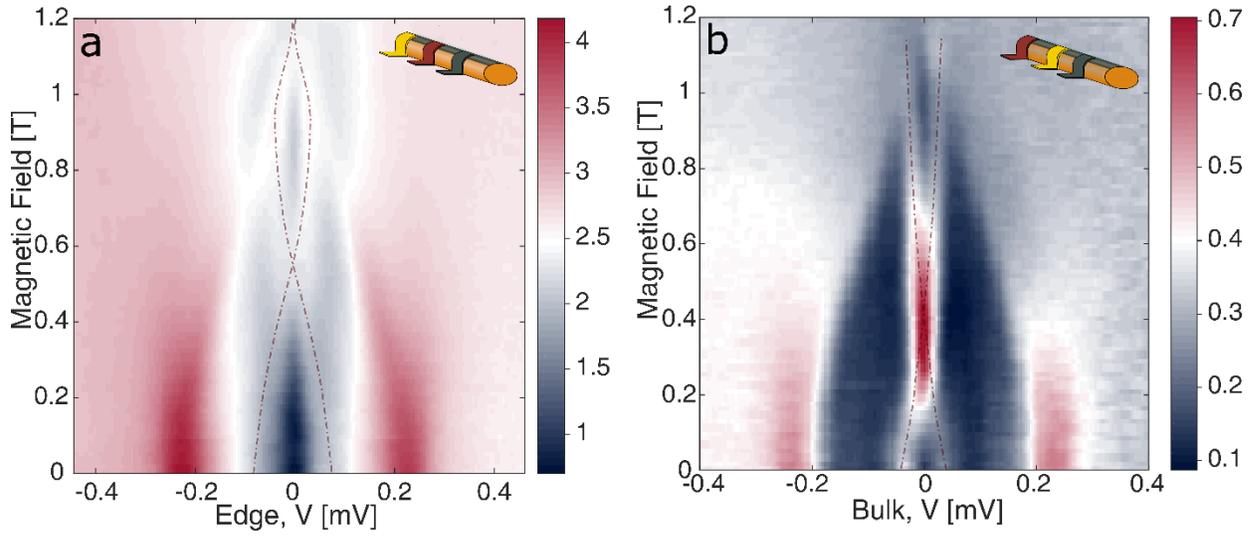

**Extended Data Figure 2. Simultaneous measurement of bulk and edge of device #2.** We repeated the experiment in another device, where we measured a gap closing at the bulk with the appearance of a ZBCP at the edge. At the edge, a ZBCP appears close to 0.5T; however, here it splits and recombines at higher field. The edge-TP has higher transparency, ~10kΩ, so that the sub-gap states features are not as sharp as in the other probes. The bulk trivial gap closes at 0.2T and reopens around 0.6T. The maximal topological gap size is ~32μV around 1T. The resistance of this bulk-TP is 73kΩ - in the range of the TPs in the reported device. This example shows that the ZBCP appears at higher field than that of the gap closing, suggesting a different chemical potential at the bulk and the edge. The back-gate voltage is $V_{BG}$=-3.0V. The segmented lines are guides to the eye.

**Measurement at a finite angle of *B* (device #3)**

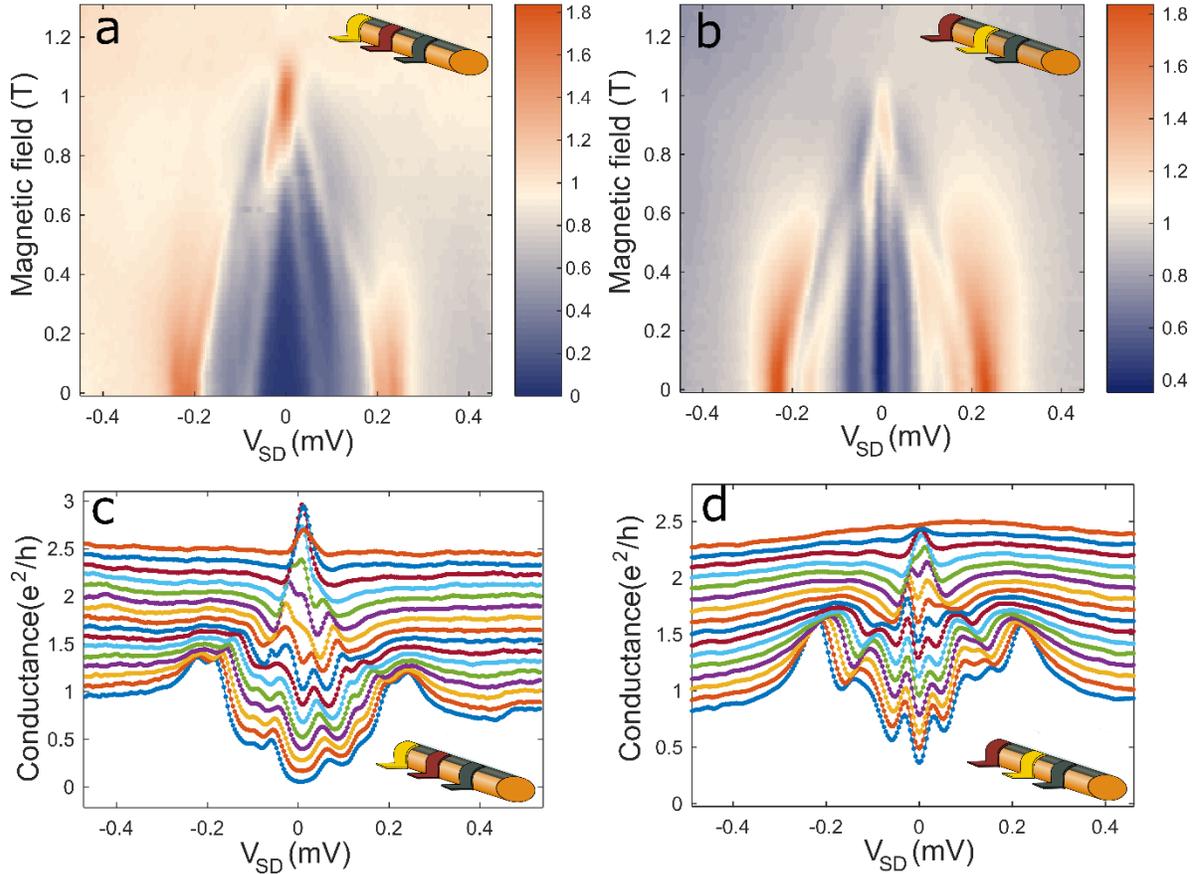

**Extended Data Figure 3. Simultaneous measurement of bulk and edge of device #3.** Repetition of the experiment in device #3, where we measured a concomitant gap closing at the bulk with the appearance of a ZBCP at the edge. In this device it was difficult to see the reopening of the bulk gap (closes at *B*~0.9T, b & d); while a strong ZBCP appear at *B*~0.9T (a & c) and remains with a conductance ~1.8$e^2/h$ (near the expected quantized value). Here, there was a finite misalignment of the nanowire with respect to the direction of the magnetic field, ~7 degrees. This resulted in a finite component of the magnetic field perpendicular to the epitaxial Al, which caused a reduction in the critical field from 1.7T to 1.2T, thus limiting the ability to see the gap reopening. The back-gate voltage is $V_{BG}$=-3.3V.

**Device after thermal recycling**

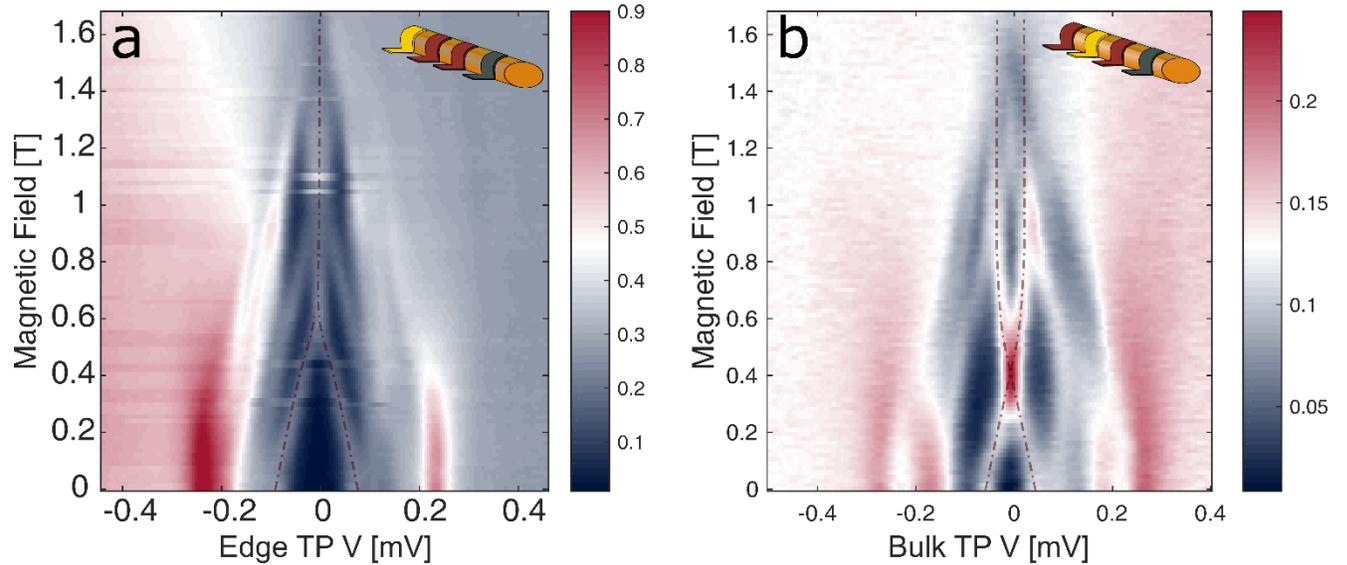

**Extended Data Figure 4. Tunneling conductance in a temperature recycled device.** (a) The edge-TP (b) The bulk-TP. a) The edge-TP shows a ZBCP, which appears at $B\sim0.6$T and survives until superconductivity diminishes at 1.6T. b) The bulk-TP shows a gap closing at $B\sim0.3$T and reopening at $B\sim0.6$T, with a maximal topological gap size of 40µeV at 0.92T. The ZBCP at the edge appears right after the gap reopens in the bulk. The back-gate voltage is $V_{BG}=-4.57$V. The segmented lines are guides to the eye.

**Non-topological ZBCP: Edge and bulk show ZBCP**

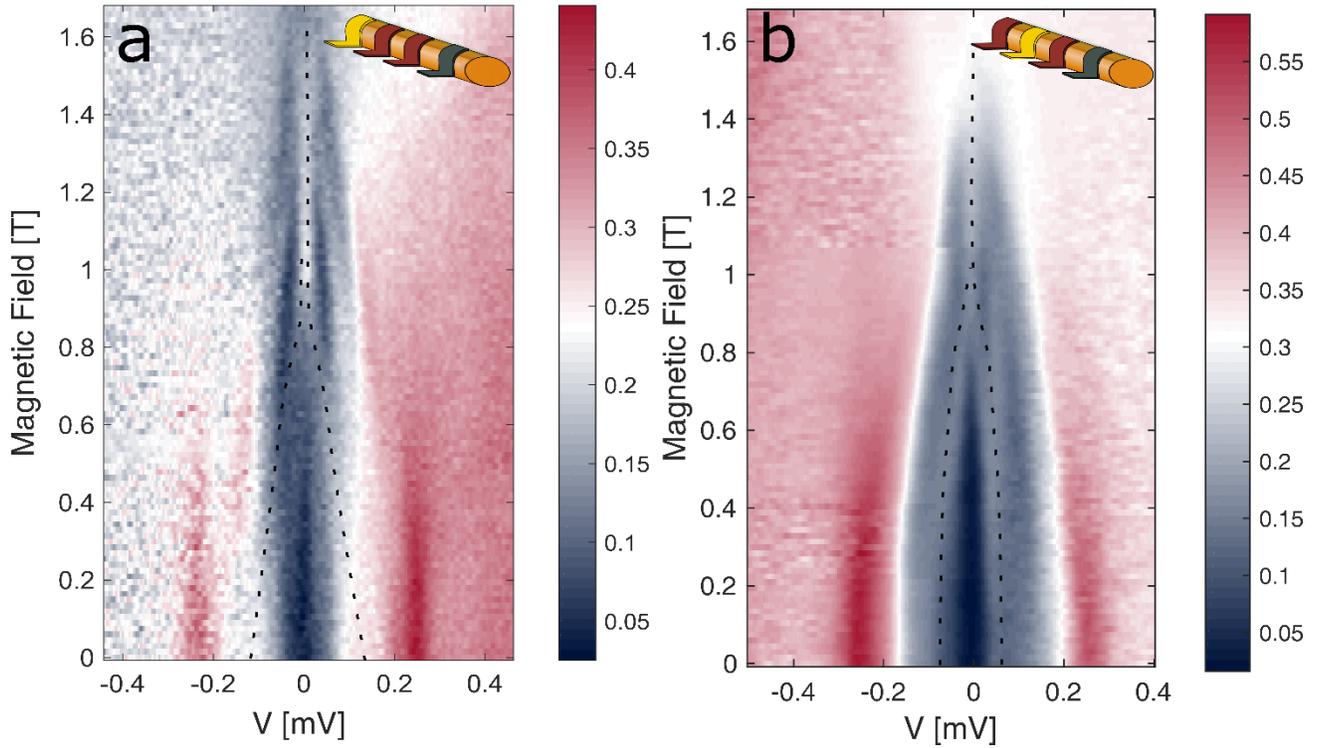

**Extended Data Figure 5. Trivial ZBCP at the edge of the nanowire.** The results reported here and in the main text are on the same device but with different back gate potential; here, $V_{BG}$=-4.29V. a) The edge-TP shows a ZBCP, which appears at $B$~1.0T and lasts until superconductivity diminished around $B$~1.7T. b) The bulk-gap closes around $B$=1.0T and doesn't reopen with magnetic field (or maybe very small and bellow our resolution); a signature of a non-topological phase in the nanowire. This is a complimentary measurement to Fig. 3 in the main text. Once the chemical potential is tuned away from the 'sweet spot', here lowering the chemical potential.

**Non-topological ZBCP: Edge and bulk show ZBCP**

In the following measurement we have changed the back-gate voltage to $V_{BG}=-4.91V$, far from the topological phase transition. A ZBCP appears around $B=0.8T$ at the edge, persisting until the superconductivity dies out. The bulk, however, shows a more complicated spectroscopic structure: a trivial zero-bias state appears and splits with applied magnetic field. A possible explanation for the absence of this state at the edge is that the wire is partly depleted, and thus the potential landscape is weakly screened, with the two tunnel-probes weakly uncorrelated.

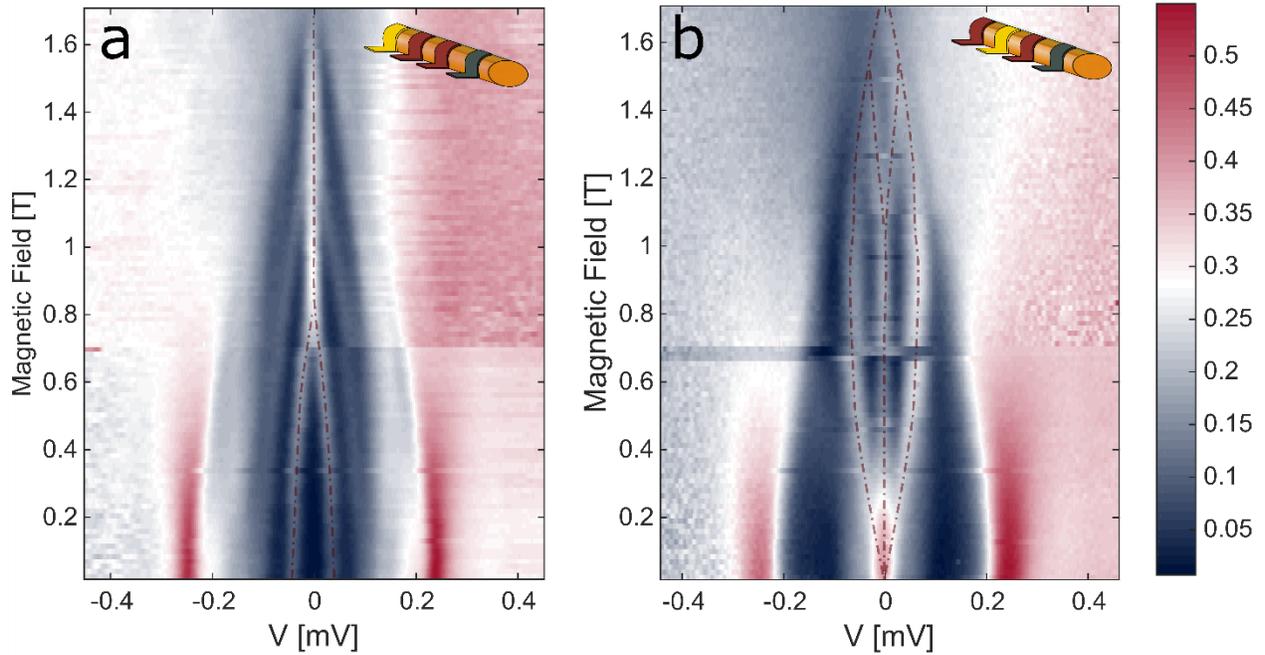

**Extended Data Figure 6. Trivial ZBCP at the edge of the wire.**
Another example of a ZBCP in the same device used in the main text, but with a trivial bulk, at $V_{BG}=-4.91V$. a) The ZBCP appears at $B\sim 0.8T$; b) The bulk-TP shows a zero-bias state at zero magnetic field. The figure also shows a $B$ dependent Andreev bound state, which splits with magnetic field in addition to another state at zero bias. The latter peak remains throughout nearly the entire range of magnetic field, splitting finally at $B\sim 1.1T$. In this configuration, even though a ZBCP emerges at the edge, the bulk is not gapped (below $B\sim 1.1T$), and the gap opens only at around $B\sim 1.1T$. This behavior is currently not understood.

**Tunnel probe characterization – bulk tunneling**

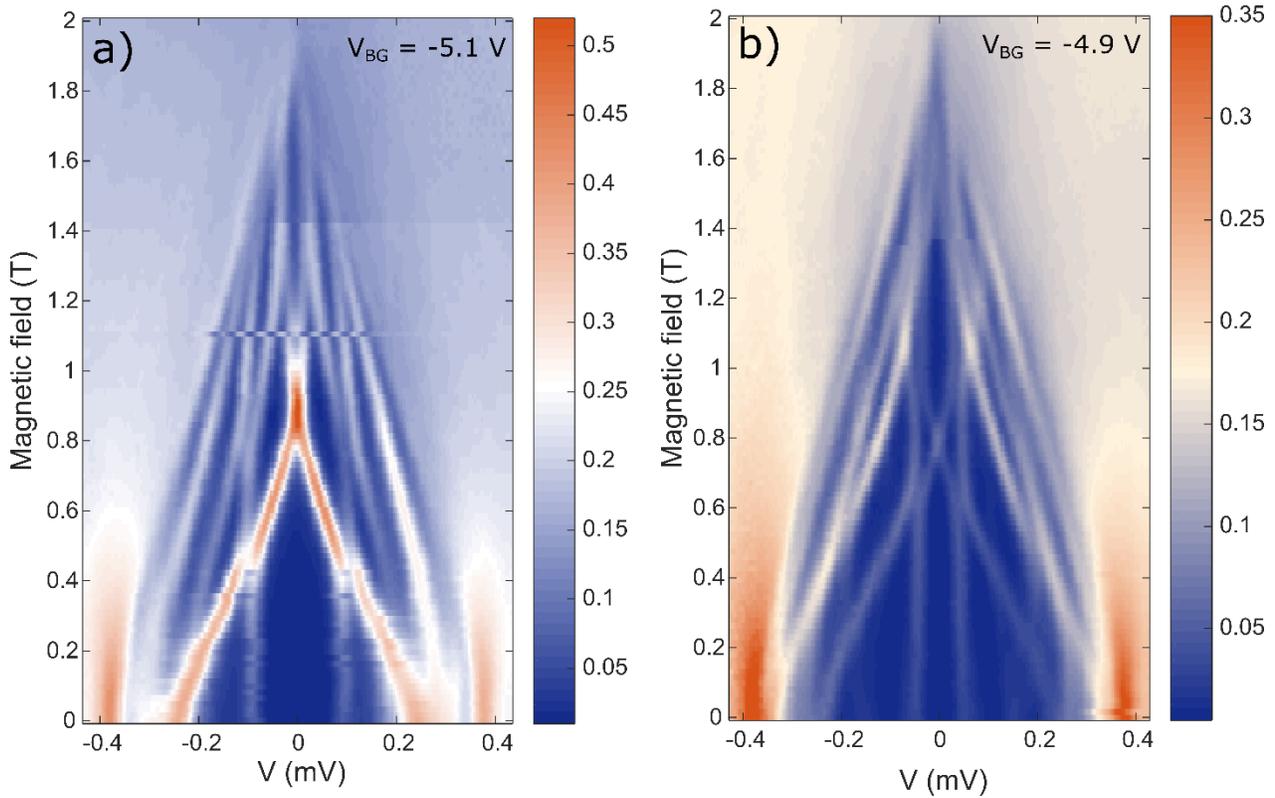

**Extended Data Figure 7. Rich structure in the bulk at low density.**

The tunnel probes we utilized for the experiment were developed especially for the purpose of studying the bulk and the edge of the InAs nanowires, aiming to have weak tunneling coupling (but not too weak). Here, we show measurements of devices with <u>only</u> with bulk-TP (resistance ~140kΩ). For a certain range of back-gate voltage a gap closer and reopening is seen. Several sub-gap states are visible and avoided crossing is observed between two pairs. a) $V_{BG}$=-5.1V, and b) $V_{BG}$=-4.9V.